\documentclass[pra,twocolumn,showpacs,preprintnumbers,amsmath,amssymb,tightenlines,superscriptaddress]{revtex4}

\usepackage{graphicx}
\usepackage{dcolumn}
\usepackage{bm}
\usepackage{epsfig}
\usepackage{stmaryrd}

 \newcommand{\intas}{\int d^{3} \mathbf{x}\:}

\newcommand{\mods}[1]{\left| #1 \right| ^{2}}
\newcommand{\deltf}{\delta^{(3)}(\mathbf{x}-\mathbf{x}')}
\newcommand{\px}{\hat{\Psi}(\mathbf{x})}
\newcommand{\padx}{\hat{\Psi}^{\dagger}(\mathbf{x})}
\def\hamax{ \hat{H}_{0a}(\mathbf{x})}
\def\hamaxs{\hat{H}_{0a}(\mathbf{x'})}

\def\phiax{\phi_{a}(\mathbf{x})}
\def\phiaxs{\phi_{a}(\mathbf{x'})}

\def\phicax{\phi^{*}_{a}(\mathbf{x})}
\def\phicaxs{\phi^{*}_{a}(\mathbf{x'})}

\def\gnx{G_{N}\left( \mathbf{x},\mathbf{x'} \right)}
\def\gnxstar{G_{N}^{*}\left( \mathbf{x},\mathbf{x'} \right)}

\def\gax{G_{A}\left( \mathbf{x},\mathbf{x'} \right)}
\def\gaxstar{G_{A}^{*}\left( \mathbf{x},\mathbf{x'} \right)}

\def\gnxx{G_{N}\left( \mathbf{x},\mathbf{x} \right)}
\def\gaxx{G_{A}\left( \mathbf{x},\mathbf{x} \right)}

\newcommand{\sub}[2]{{#1}_{\mbox{\!\! \scriptsize #2}}}
\newcommand{\subsup}[3]{{#1}_{\mbox{\!\! \scriptsize #2}}^{\mbox{\!\! \scriptsize #3}} }

\newcommand{\bv}[1]{\mathbf{ #1 }}

\def\noi{\noindent}
\def\beq{\begin{align}}
\def\eeq{\end{align}}
\def\nnl{\\[0.15cm] \nonumber}
\def\nl{\\[0.15cm] }
\def\CR{\nonumber\\[0.15cm]}

\newcommand{\fref}[1]{Fig.~\ref{#1}}
\newcommand{\eref}[1]{Eq.~(\ref{#1})}
\newcommand{\sref}[1]{section~\ref{#1}}
\newcommand{\aref}[1]{appendix~\ref{#1}}

\begin{document}

\title{Quantum depletion of collapsing Bose-Einstein condensates}
\author{Sebastian~W\"uster}
\affiliation{ARC Centre of Excellence for Quantum-Atom Optics, Department of Physics, Faculty of Science, Australian National University, Canberra ACT 0200, Australia}
\author{Beata~J.~D\c{a}browska-W\"uster}
\affiliation{ARC Centre of Excellence for Quantum-Atom Optics, 
Nonlinear Physics Centre, Research School of Physical Sciences and Engineering, Australian National University, 
Canberra ACT 0200, Australia}
\author{Ashton~S.~Bradley}
\affiliation{ARC Centre of Excellence for Quantum-Atom Optics, School of Physical Sciences, University of Queensland, Brisbane QLD 4072, Australia}
\author{Matthew~J.~Davis}
\affiliation{ARC Centre of Excellence for Quantum-Atom Optics, School of Physical Sciences, University of Queensland, Brisbane QLD 4072, Australia}
\author{P.~Blair~Blakie}
\affiliation{Physics Department, University of Otago, P.O. Box 56, Dunedin, New Zealand }
\author{Joseph.~J.~Hope}
\affiliation{ARC Centre of Excellence for Quantum-Atom Optics, Department of Physics, Faculty of Science, Australian National University, Canberra ACT 0200, Australia}
\author{Craig.~M.~Savage}
\affiliation{ARC Centre of Excellence for Quantum-Atom Optics, Department of Physics, Faculty of Science, Australian National University, Canberra ACT 0200, Australia}
\email{craig.savage@anu.edu.au}

\begin{abstract}
We perform the first numerical three-dimensional studies of quantum field effects in the Bosenova experiment on collapsing condensates by E.~Donley {\it et al.} [Nature {\bf415}, 39 (2002)] using the exact experimental geometry. In a stochastic truncated Wigner simulation of the collapse, the collapse times are larger than the experimentally measured values. We find that a finite temperature initial state leads to an increased creation rate of uncondensed atoms, but not to a reduction of the collapse time. A comparison of the time-dependent Hartree-Fock-Bogoliubov and Wigner methods for the more tractable spherical trap shows excellent agreement between the uncondensed populations. We conclude that the discrepancy between the experimental and theoretical values of the collapse time cannot be explained by Gaussian quantum fluctuations or finite temperature effects.
\end{abstract}


\maketitle

\section{Introduction}
\label{intro}
Experimental progress in dilute gas Bose-Einstein condensates (BECs) has recently 
allowed increasingly detailed studies of the quantum nature of the atomic field~\cite{Schellekens2005,Chuu2005,Folling2005,Ottl2005}.
This has been accompanied by advances in the numerical treatment of many-body quantum field theory applied to BEC dynamics, most notably in a better understanding of phase space methods~\cite{book:qn} and Hartree-Fock-Bogoliubov theory~\cite{morgan:hfb}.
Therefore, experiments in which the physics is sufficiently straight-forward that quantitative agreement with many-body quantum theory can be expected are especially appealing. 

In this paper we extend our previous analyses~\cite{savage:coll,wuester:nova} of one of these: the JILA {\it Bosenova} experiment of E.~Donley {\it et al.}~\cite{jila:nova}, in which $^{85}$Rb BECs were made to collapse by switching their atomic interactions to attractive. Many interesting phenomena associated with the collapse, like bursts and jets, have attracted widespread attention. These have been understood qualitatively using a variety of models~\cite{adhikari:coll,adhikari:jets,adhikari:lattice,saito_ueda:coll,saito_ueda:intermitt,saito_ueda:powerlaw,saito_ueda:picture,shlyapnikov:coll,savage:coll,bao:jets,duinestoof:feshbachdynamics,duinestoof:evaporation,metens,calz:hu,holland:burst,wuester:nova}.
However precise \emph{quantitative} agreement, of the kind sought in this paper, has not been fully achieved. 

Here we are concerned with a quantitative description of the most basic aspect of the collapse experiment: the time to initiation of the collapse, $\sub{t}{collapse}$. The abrupt onset of atom losses in the experiment allows a precise measurement of this time. It has previously been shown that the Gross-Pitaevskii (GP) theory substantially overestimates these collapse times~\cite{savage:coll}. However quantum corrections in the framework of time-dependent Hartree-Fock-Bogoliubov (HFB) theory were shown not to accelerate the collapse of a BEC in a spherically symmetric trap~\cite{wuester:nova}. Here we investigate the collapse in a cigar shaped trap, exactly as in the experiment. Our simulations use the stochastic truncated Wigner approximation (TWA), with the experimental parameters. We find that the inclusion of quantum effects does not yield results in agreement with the experiment. Therefore, for the Bosenova experiment, despite an excellent qualitative understanding, we do not yet have quantitatively precise theoretical models for even the simplest aspects. The implications of this are discussed in the conclusion.

We also compare the TWA with the HFB formalism for a spherically symmetric trap. We find that the quantum depletion predicted by both methods agrees very well. As both methods independently confirm an excited state population insufficient to accelerate the collapse, we can rule out zero temperature depletion as a mechanism for collapse acceleration. 

Moreover, using both quantum field methods we show that the initial presence of a thermal cloud increases the production rate of uncondensed atoms. This results in a reduction of the condensate population just before collapse, which in principle could appear as a slightly accelerated collapse. However, as we explain in \sref{thermal}, this effect would be difficult to detect in an experiment. Irrespective of this, even for temperatures about three times higher than the experimentally measured temperature of $T=3$ nK, the acceleration of the collapse due to depletion is insufficient to bring the theoretical and experimental results into agreement. 

In summary, we present a careful quantitative study of the best characterized experimental data in the Bosenova experiment: the collapse time. We find that not only the GP model, but also the HFB and TWA theories fail to explain the collapse times. 
Further experimental and theoretical work should resolve this unsatisfactory situation.

This paper is organized as follows: In \sref{experiment} we give a brief overview of the Bosenova experiment. Section \ref{models} contains the theoretical background of the TWA and HFB methods for the quantum field description of BECs. This is followed in \sref{numerics} by a discussion of the numerical limitations of the TWA. Sections \ref{cigar} and \ref{thermal} contain the results of the TWA for the collapse of a BEC in a cigar trap. Finally in \sref{spherical} we report on the comparison between HFB and TWA in a spherically symmetric case.

\section{The Bosenova experiment}
\label{experiment}
In the experiment~\cite{jila:nova}, a stable $^{85}$Rb condensate was prepared with scattering length $a_{s}=0$ using a Feshbach resonance, before $a_{s}$ was switched to a negative (attractive) value $a_{s}=\sub{a}{collapse}$. The resulting collapsing condensate was observed to lose atoms until the atom number was reduced to about the critical value below which a stable condensate can exist~\cite{jila:nova}. Usually the remnant atom number was found to be slightly greater than the critical value, a puzzle which has only recently been explained, with the help of a new experiment~\cite{jila:solitons}, by the formation of multiple bright solitons. 

The onset of atom number reduction is quite sudden. After the change in scattering length a few milliseconds of very little loss is observed. This is followed by a rapid decay of condensate population (within $\sim0.5$ ms) after which the condensate stabilizes again. This behavior results from the scaling of the loss rate with the cube of the density, the peak value of which rises as $1/(\sub{t}{collapse}-t)$ near the collapse point~\cite{shlyapnikov:coll}. The sudden onset of atom loss allows a precise definition of the collapse time $\sub{t}{collapse}$, as the time after initiation of the collapse ($a_{s}\rightarrow \sub{a}{collapse}$) up to which atom loss remains negligible. In this paper we focus our attention primarily on a case with $\sub{a}{collapse}=-10\sub{a}{0}$, where $\sub{a}{0}$ is the Bohr radius. For this case the experimentally measured $\sub{t}{collapse}$ is $(6\pm1)$ ms~\cite{jila:nova,jila:revision}, while Gross-Pitaevskii studies found it to be about $10$ ms~\cite{savage:coll}.

A quantitative result of the experiment is the dependence of $\sub{t}{collapse}$ on the magnitude of the attractive interaction, parametrized by the (negative) scattering length $\sub{a}{collapse}$. These measurements are performed starting with $\sub{N}{ini}=6000$ atoms in an ideal gas state, i.e.~the interaction between atoms is tuned to zero. 
The $\sub{t}{collapse}$ data points presented in~\cite{jila:nova} have undergone one revision of their $\sub{a}{collapse}$ values by a factor of 1.166(8) due to a more precisely determined background scattering length~\cite{jila:revision}. 

Other experimental features like the bursts and jets mentioned in the introduction have also been measured in great detail, tempting quantitative explanation. However the HFB and TWA methods used in this paper become impractical soon after the initiation of collapse, and are therefore unsuitable for analysis of the full collapse, even if they correctly modelled the collapse times. Refs.~\cite{wuester:nova,saito_ueda:picture} review the state of theoretical studies of the Bosenova experiment.

\section{Quantum field Models of a harmonically trapped BEC \label{models}}
The Hamiltonian for a Bose gas of interacting atoms in an external trapping potential $V(\mathbf{x})$ is given by:
\begin{align}
\hat{H}&=\intas \,\padx \hat{H}_{0}\px
\CR
&+\frac{\sub{U}{0}}{2}\intas \,\padx\padx \px\px,
\label{hamiltonian}
\\
\hat{H}_{0}&=-\frac{\hbar^2}{2 m} \nabla^2 + V(\mathbf{x}).
\nonumber
\end{align}
Here $\px$ ($\padx$) is the field operator that annihilates (creates) a boson at position $\mathbf{x}$, 
$m$ is the atomic mass and $\sub{U}{0}=4 \pi \hbar^2 a_s/m$ is the interaction strength with 
the $s$-wave scattering length $a_s$. In the following we use parameters corresponding to the collapse experiment~\cite{jila:nova}. The $^{85}$Rb atoms with $m=1.41 \times 10^{-25}$ kg are confined in a cigar shaped cylindrically symmetric trap $V(\mathbf{x})=m(\omega_{\perp}^{2}r^{2} + \omega_{z}^{2}z^{2})/2$, where the trap frequencies are $\omega_{\perp}=17.5\times2\pi$~Hz and $\omega_{z}=6.8\times2\pi$~Hz. 
For comparative purposes, we additionally consider a spherical trap with the geometric mean frequency $\bar{\omega}=(\omega_{\perp}^{2}\omega_{z})^{1/3}=12.8\times2\pi$~Hz in \sref{spherical}. 

In the actual BEC collapse atom losses due to three-body recombination play a crucial role. These losses are taken into account in the master equation for the time evolution of the system's density operator $\hat{\rho}$~\cite{jack:loss}:
\begin{align}
\frac{d \hat{\rho}}{d t}&=-\frac{i}{\hbar}[\hat{\rho},\hat{H}]
+\frac{K_{3}}{6}\intas \big( 2\px^{3}\rho \padx^{3}
\CR
&-\padx^{3}\px^{3}\rho-\rho\padx^{3}\px^{3} \big).
\label{mastereqn}
\end{align}
The three-body loss constant $K_3=1\times 10^{-39}\mbox{m}^{6}\mbox{s}^{-1}$ is chosen as in~\cite{savage:coll}. $K_{3}$ is not very well constrained experimentally, but in simulations its value can be varied by factors of 10 to 100 without significantly affecting the collapse time~\cite{savage:coll}. The reason for this is that the three-body loss acts only as a diagnostic for a rapid increase in density at the point of collapse. In the time leading to this increase, the density of the contracting BEC remains low enough for three-body loss to play no role in the dynamics. Only at and after the actual time of collapse does the precise value of $K_{3}$ become relevant. 

In the following subsections we describe two different approaches to finding approximate solutions of the quantum evolution given by \eref{mastereqn}.

\subsection{Truncated Wigner method}
\label{wigner}
To obtain the time evolution of $\hat{\rho}$ we may represent $\hat{\rho}$ in a suitable phase-space~\cite{book:qn}. In this case we make use of the Wigner representation.
We define the multimode Wigner function:
\begin{align}
W(\{\alpha_{k}\},\{\alpha_{k}^*\})&=\prod_{k} \left(\int \frac{1}{\pi^{2}}d^2\beta_{k}\right)
\exp{\Big(\sum_{k}\beta_{k}^*\alpha_{k}-\beta_{k}\alpha_{k}^* \Big)}
\CR
& \mbox{Tr} \Big[\exp{\Big(\sum_{j}\beta_{j}\hat{a}^{\dag}_{j} - \beta_{j}^*\hat{a}_{j}} \Big)\hat{\rho} \Big].
\label{wignermultimode}
\end{align}
Here $\hat{a}_{j}^{\dag}$ ($\hat{a}_{j}$) creates (annihilates) atoms in the $j$th single particle mode. 
These may be eigenstates of the harmonic oscillator or position eigenstates on a discrete grid. Using this representation it is possible to obtain the evolution of $W(\{\alpha_{k}\},\{\alpha_{k}^*\})$ from \eref{mastereqn}. If the resulting equation is \emph{truncated} by neglecting derivatives higher than second order with respect to the $\alpha_{k}$, it takes the form of a Fokker-Planck equation (FPE), which can then be mapped onto a stochastic differential equation (SDE)~\cite{book:qn}. 
The validity criteria for the truncation and details of the procedure are discussed in Refs.~\cite{steel:wigner,castin:validity,norrie:long}. 

Other theoretical approaches, such as the positive-P or gauge-P phase space methods~\cite{drummond:posp,drummond:gaugep}, can include the full quantum
evolution of the s-wave scattering physics, but still necessitate an approximate description of three-body losses. We have implemented both these methods for a one dimensional, and also for a spherically symmetric, collapse scenario and found them to be numerically intractable.

Compared to the situation without loss~\cite{steel:wigner} the inclusion of three-body recombination in the master equation results in additional terms in the stochastic differential equation. These have been thoroughly treated in~\cite{ashton:loss}. Drawing on these previous studies, we can write down the \emph{simplest} SDE~\cite{foonote:simplest} which describes a trapped BEC with three-body loss: 
%
\begin{align}
d \phi(\mathbf{x})
&=-\frac{i}{\hbar}\left(-\frac{\hbar^2}{2 m} \nabla^2 + V(\mathbf{x}) + \sub{U}{0}|\phi(\mathbf{x})|^2\right)\! \phi(\mathbf{x})dt
\CR
&-\frac{K_3}{2}|\phi(\mathbf{x})|^4 \phi(\mathbf{x})dt +\sqrt{\frac{3 K_3}{2}}|\phi(\mathbf{x})|^2 d\xi(x,t).
\label{sgpe}
\end{align}
Details regarding the construction of the dynamical noise term $d\xi(x,t)$ are given in \aref{pgpe_appendix}. 
By setting $d\xi(x,t)=0$ in \eref{sgpe} we can recover the usual Gross-Pitaevskii equation including three-body loss~\cite{savage:coll}.  
\par
The initial state of the stochastic wavefunction $\phi(\bv{x})$ has to be chosen such that it represents the Wigner function of an initial coherent state BEC. At zero temperature this is achieved by the addition of initial vacuum noise $\eta(\bv{x})$:
\begin{align}
\phi(\bv{x},t=0)&=\psi_{0}(\bv{x}) + \frac{1}{\sqrt{2}}\eta(\bv{x}).
\label{ininoise}
\end{align}
Here $\psi_{0}(\bv{x})$ denotes the oscillator ground state, which is appropriate since the starting point of the experiment is a non-interacting BEC. $\eta(\bv{x})$ is a Gaussian distributed complex random function that fulfills the conditions $\overline{\eta(\bv{x})\eta(\bv{x'})} = 0$ and $\overline{ \eta(\bv{x})^{*}\eta(\bv{x'})} = \delta(\bv{x}-\bv{x'})$, where $\overline{f}$ denotes the stochastic average of $f$.

The truncation of higher order derivatives in the FPE is safely applicable when all modes in the problem are highly occupied~\cite{castin:validity}. If the three-dimensional collapse scenario is numerically solved on a spatial grid as in~\cite{savage:coll}, 6000 atoms are spread out over $\sim 4 \times 10^{6}$ position space modes (i.e. grid points). Thus in the position basis the mode occupation criterion cannot be fulfilled. Also the addition of the initial noise as in \eref{ininoise} becomes problematic, leading to aliasing effects at the edge of the computational grid. These can be overcome in periodic situations as described in~\cite{beatka:train}, but  would be persistent in a harmonic trap. 

Here we choose a powerful method to solve the Gross-Pitaevskii equation in the oscillator (energy) basis instead. The method was presented in~\cite{dion:hermites} and applied to BECs at finite temperature in~\cite{matthewandblair:pgpe,matthewandblair:pgpeprl}. 
In our work we extend the formalism in order to solve \eref{sgpe}. The stochastic field $\phi(\bv{x})$ in \eref{sgpe} can be expanded in terms of eigenstates $\varphi$ of the 1D harmonic oscillator:
\begin{align}
\phi(\mathbf{x},t)&=\sum_{ \{l,m,n\} \in {\cal C} }c_{lmn}(t)\varphi_{l}(x)\varphi_{m}(y)\varphi_{n}(z),
\label{osciexpansion}
\end{align}
which fulfill:
\begin{align}
\left(-\frac{\hbar^2}{2 m}\frac{\partial^{2}}{\partial x^{2}} + \frac{1}{2}m\omega_{x}^{2}x^{2} \right)\varphi_{l}(x)&=\epsilon_{l}\varphi_{l}(x).
\end{align}
Similarly $\varphi_{m}(y)$ and $\varphi_{n}(z)$ solve the oscillator equation in the $y$- and $z$-dimensions. The summation in \eref{osciexpansion} is restricted to all modes with total energy below a certain cutoff  $\sub{E}{cut}$:
\begin{align}
{\cal C}=\left\{l,m,n: \:\: \epsilon_{l} +\epsilon_{m}+\epsilon_{n}  \leq \sub{E}{cut}\right\}. 
\end{align}
The values for $\sub{E}{cut}$ are given in units of $\hbar \omega_{\perp}$ ($\hbar \bar{\omega}$) for the cylindrical (spherical) case. 
Further details of the approach are given in \aref{pgpe_appendix}.

The stochastic equation (\ref{sgpe}) has to be solved for many realizations (trajectories)~\cite{footnote:trajectories}. If the Wigner representation is used, symmetrized quantum averages can then be determined from averages over all trajectories~\cite{book:qn}. In what follows $\langle \hat{f} \rangle$ denotes the quantum expectation value of $\hat{f}$. We obtain the condensed (coherent) and uncondensed (incoherent) populations for each oscillator mode from:
\begin{align}
\subsup{n}{cond}{$lmn$}&=\left| \langle \hat{\Psi}_{lmn} \rangle \right|^{2}=\left| \overline{c_{lmn}} \right|^{2},
\label{ncond}
\\
\subsup{n}{unc}{$lmn$}&=\langle \hat{\Psi}^{\dagger}_{lmn} \hat{\Psi}_{lmn} \rangle - \left| \langle \hat{\Psi}_{lmn} \rangle \right|^{2}
\CR
&=\overline{|c_{lmn}|^{2}} - \subsup{n}{cond}{$lmn$} - \frac{1}{2}.
\label{}
\end{align}
The $\hat{\Psi}_{lmn}$ ($\hat{\Psi}_{lmn}^{\dagger}$) are field operators that annihilate (create) an atom in the mode with quantum numbers $l,m,n$. The numbers of condensed and uncondensed atoms ($\sub{N}{cond}$, $\sub{N}{unc}$) are then obtained by summing the populations in all the modes. The total atom number is given by $\sub{N}{tot}=\sub{N}{cond}+\sub{N}{unc}$.

A more rigorous definition of the condensate component of the stochastic field is given by the Penrose-Onsager criterion~\cite{penrose:condensate}. Exemplary applications of this method can be found in~\cite{matthewandblair:pgpe,gajda:classfield}. To employ the criterion we would need to average and subsequently diagonalize the one-body density matrix of size $\sub{N}{modes}\times\sub{N}{modes}$, which however is not feasible in our case as will be explained in \sref{numerics}.

Finally we point out that the mode occupation criterion of~\cite{castin:validity} can be slightly relaxed to the requirement that the noise density defined by $\delta_{c}(\bv{x},\bv{x})\equiv \sum_{\{l,m,n\} \in {\cal C}}\left|\varphi_{l}(x)\varphi_{m}(y)\varphi_{n}(z) \right|^{2}$ is smaller than the condensate density $n_{c}$ within the volume where the latter is significant~\cite{norrie:prl,norrie:long}. This criterion is basis dependent and for the Bosenova problem it can be fulfilled in the oscillator basis but not in the position basis.

\subsection{Time-dependent Hartree-Fock-Bogoliubov approach}
\label{hfb}
A different method to go beyond the mean-field theory is to derive the Heisenberg equation for the field operator $\hat{\Psi}_{a}(\mathbf{x},t)$, and subsequently decompose $\hat{\Psi}_{a}(\mathbf{x},t)$ into a condensate part $\phi_{a} (\mathbf{x},t)$ and quantum fluctuations $\hat{\chi}(\mathbf{x},t)$, such that $\hat{\Psi}_{a}=\phi_{a}+\hat{\chi}$ and $\langle \hat{\Psi}_{a} \rangle=\phi_{a}$. The quantum fluctuations can be described in terms of their lowest order correlation functions: the normal density $G_{N}(\mathbf{x},\mathbf{x}') = \langle \hat{\chi}^{\dagger} (\mathbf{x}')\hat{\chi}(\mathbf{x}) \rangle$ and anomalous density $G_{A}(\mathbf{x},\mathbf{x}') = \langle \hat{\chi}(\mathbf{x}') \hat{\chi}(\mathbf{x}) \rangle$. The derivation of the dynamical equation for the condensate contains a factorization of the expectation values in accordance with Wick's theorem~\cite{book:blaizot}, 
e.g.~$\langle \hat{\Psi}_{a}^{\dagger}\hat{\Psi}_{a}\hat{\Psi}_{a}\rangle=
2\langle \hat{\Psi}_{a}^{\dagger}\hat{\Psi}_{a}\rangle \langle\hat{\Psi}_{a}\rangle
+\langle \hat{\Psi}_{a}^{\dagger}\rangle \langle\hat{\Psi}_{a}\hat{\Psi}_{a}\rangle$. This implies the assumption that the system is in a Gaussian quantum state (i.e.~a coherent state or even a squeezed state). We obtain as dynamical equation for the condensate:
\begin{align}
i \hbar  &\frac{\partial \phiax}{\partial t}=\left(-\frac{\hbar^2}{2m} \nabla^2 +V(\mathbf{x}) 
 +\sub{U}{0} |\phiax |^{2} \right)  \phiax
\CR
&+ 2 \sub{U}{0} \phiax \gnxx +\sub{U}{0}\phicax \gaxx.
\label{atom_eqn}
\end{align}
Here $\gnxx$ represents the density of uncondensed atoms. Hence this modified Gross-Pitaevskii equation contains the interaction of the uncondensed component with the mean-field. 

We also phenomenologically model three-body loss from the condensate, by adding the following term to equation (\ref{atom_eqn}):
\begin{align}
&-i \frac{\hbar}{2} K_{3}  |\phiax |^{4}\phiax.
\label{lossterm}
\end{align} 
To obtain the time evolution of the condensate we have to supplement \eref{atom_eqn} by evolution equations for $G_{N}$ and $G_{A}$, which are listed in \aref{hfb_appendix}.
We refer to Ref.~\cite{wuester:nova} for further details regarding a method to solve the set of coupled equations in a harmonic trap. These equations require a renormalisation of the coupling strength $U_{0} $ due to the momentum cutoff $K=\pi/\Delta x$ of the numerical grid used in the simulations~\cite{holland:renorm}. One must distinguish the physical interaction strength $U$ and the parameter $\sub{U}{0}$ in the Hamiltonian, which are related by: 
\begin{align}
\sub{U}{0}&=\dfrac{U}{1-\alpha \;U},\:\:\:\:
\alpha=\dfrac{m K}{2 \pi^{2} \hbar^{2}} .
\label{renormalization}
\end{align} 
A similar renormalisation issue arises in the truncated Wigner method where the same prescription has to be used to relate the numerical coupling to the physical interaction strength~\cite{castin:validity}. In the interaction strength regime of interest for this paper the difference between $U$ and $\sub{U}{0}$ is negligible.  A careful renormalisation is hence unnecessary and we can directly employ $\sub{U}{0}=U=4 \pi \hbar^2 a_s/m$ in our simulations. 

The cutoff is determined by the spatial lattice spacing in the HFB case, and by the energy cutoff $\sub{E}{cut}$ in the TWA case, after equating the free particle kinetic energy to the energy cutoff: $\sub{E}{cut}=\hbar^{2}K^{2}/(2m)$. The cutoffs are chosen to ensure numerical accuracy of the simulations, and in particular that the results of interest are invariant with respect to changes in the cutoffs.
For our HFB simulations the highest cutoff is $K = 1.3 \times 10^7$ m$^{-1}$, and $|\alpha U| = 4.5 \times 10^{-3}$, corresponding to less than one-half percent renormalization.
For our TWA simulations the maximum cutoff is $\sub{E}{cut} \approx 50$ and therefore $K = 3.8 \times 10^6$ m$^{-1}$, and $|\alpha U| = 1.3 \times 10^{-3}$, again corresponding to negligible renormalization.

We note that the three-body loss term in the HFB formalism, \eref{lossterm}, only incorporates loss processes among condensate particles, whereas the implementation in \eref{sgpe} of the TWA contains loss also for the uncondensed fraction. The comparison between the methods in this paper is done for very small uncondensed population and small total losses, so that this difference can be neglected.

\section{Numerical constraints}
\label{numerics}
The aim of this section is to clarify the basis of the conclusions drawn from our simulations. 

In this article we present solutions of \eref{sgpe}, modeling the Bosenova experiment without any significant free parameters, using three different levels of approximation of the truncated Wigner method:
\begin{itemize}
\item{GP evolution: Gross-Pitaevskii evolution only.  In this case both the noise on the initial state and dynamical noise are omitted ($\eta=0$, $d\xi=0$).}
\item{TWA with initial noise: Truncated Wigner evolution without dynamical noise ($d\xi=0$).}
\item{TWA with dynamical noise: Complete truncated Wigner evolution ($\eta\neq0$, $d\xi\neq0$).}
\end{itemize}

The reasons for studying the GP evolution are two-fold. Firstly it aids the determination of  the required number of oscillator modes by comparison with the established position space results~\cite{savage:coll}. Secondly it allows us to quantify the differences between the classical and quantum field results. 
\begin{figure}[htb]
\centering
\epsfig{file={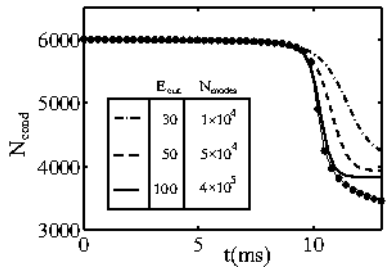},width=6.5cm} 
\caption{GP evolution only. Atom number $\sub{N}{cond}$ in the condensate during collapse with $\sub{a}{collapse}=-10 \sub{a}{0}$. ($\bullet$) Results obtained on a spatial grid~\cite{savage:coll}. Thick lines correspond to solutions of \eref{sgpe} in the energy basis. Inset: A table with the number of modes for different $\sub{E}{cut}$ and a legend. As $\sub{E}{cut}$ increases, the atom number curves approach the correct position basis solution.
}
\label{gponly_comparemodes}
\end{figure}

To determine the required mode numbers, the GP equation is solved in the harmonic oscillator basis in order to reproduce the atom number curve of Ref.~\cite{savage:coll}. In doing so, we encountered a limitation of the oscillator-basis: due to the extremely narrow peak of the condensate wavefunction at the collapse time~\cite{saito_ueda:coll}, numerically accurate simulations beyond this point require a very large number of modes, $\gtrsim 10^{6}$. 
Simulating the condensate evolution much beyond the collapse time is therefore not feasible~\cite{footnote:notfft}.  
\fref{gponly_comparemodes} shows the number of atoms remaining in the condensate for different numbers of oscillator modes employed. In the case of $4 \times 10^{5}$ modes, the result appears close to convergence against the  solution of the GP obtained on a spatial grid~\cite{savage:coll}. However, we can conclude from the evolution of the peak densities that a cutoff of at least $\sub{E}{cut}\gtrsim150$, corresponding to about $1.5\times 10^{6}$ modes, would be required to evolve through the collapse. Nevertheless, we find that the evolution until $\sim\!8$ ms can be accurately represented with a basis-size that is computationally tractable in the stochastic multi-trajectory treatment ($\sim 5\times 10^{4}$ modes, $\sub{E}{cut}=50$). With this mode-number, the validity criterion of~\cite{norrie:long} is safely fulfilled.

\section{Collapse of a cigar shaped BEC}
\label{cigar}
If \eref{sgpe} is solved in the truncated Wigner formalism, the condensed and the uncondensed fractions of the atomic gas can be distinguished. During the collapse population is transferred between the fractions due to the interaction. Since the interaction between the condensed and uncondensed components is twice as strong
as the self-interaction of the condensate (compare \eref{atom_eqn}), a sufficiently strong quantum depletion could possibly yield a quicker collapse than a pure BEC. 

\begin{figure}[htb]
\centering
\epsfig{file={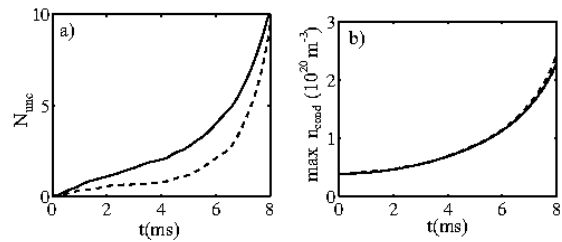},width=\columnwidth} 
\caption{Initial $8$ ms of evolution after the change in scattering length from $0$ to $\sub{a}{collapse}=-10 \sub{a}{0}$. (a) $\sub{N}{unc}$ for the solutions with dynamical noise (solid) and initial noise only (dashed). (b) Condensate peak density for GP evolution (solid) and TWA with dynamical noise (dashed). The result is unchanged, hence no acceleration of the collapse occurred before $8$ ms, which exceeds the experimental collapse time of  $(6\pm1)$ ms. In both (a) and (b) $\sub{E}{cut}=50$.}
\label{wigner_collapse}
\end{figure}

Here we present results for a collapse with $\sub{a}{collapse}=-10\sub{a}{0}$, for which the experimentally measured $\sub{t}{collapse}$ is $(6\pm1)$ ms~\cite{jila:nova,jila:revision}. We find that only very few uncondensed atoms are created prior to the actual collapse, see \fref{wigner_collapse}~(a). This result qualitatively agrees with our previous studies of a spherically symmetric geometry with the HFB method~\cite{wuester:nova}. Due to the very small depletion in the initial stage, our quantum treatment does not show an acceleration of the collapse compared to the GP evolution. We deduce this from the identical peak densities in \fref{wigner_collapse}~(b). As discussed in the previous section, despite the numerical limitations we can thus conclude that the inclusion of zero temperature quantum depletion does not result in agreement between theory and experiment. 

\fref{wigner_collapse}~(a) shows that the dynamical noise contributes significantly to the evolution of the uncondensed atom number. \fref{wigner_hfb}~(b) confirms that dynamical noise is necessary for exact agreement with the HFB approach.

We have checked that these results are qualitatively unchanged for other scenarios, i.e.~where the attractive interaction is changed to $\sub{a}{collapse}=-6\sub{a}{0}$ and $\sub{a}{collapse}=-25\sub{a}{0}$. In the latter case, we have observed a larger production of uncondensed atoms just before collapse, which also did not significantly accelerate the collapse. 

\section{Inclusion of a thermal cloud\label{thermal}}
As a next step towards even more accurate modelling of the experiment we include initial thermal population in the uncondensed modes. We will show that, if temperature effects are taken into account, the precise results for the measured collapse times might depend on whether $\sub{N}{cond}$ or $\sub{N}{tot}$ is measured.

\begin{figure}[htb]
\centering
\epsfig{file={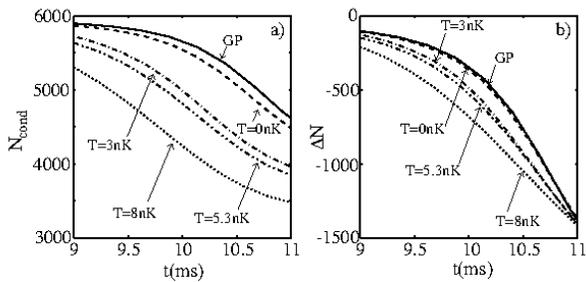},width=\columnwidth} 
\caption{Slight acceleration of the collapse due to initial thermal atoms. (a) Time evolution of $\sub{N}{cond}$ around the collapse.  
(b) Change in total number $\Delta N=\sub{N}{tot}(t)-\sub{N}{tot}(0)$ for the same period of the evolution. The sampling errors of all these results are less than $1\%$~\cite{footnote:samplingerror}. $\sub{E}{cut}=50$.}
\label{thermal_collapse}
\end{figure}

In the oscillator basis representation, the initial state for nonzero temperature~\cite{castin:validity} is
\begin{align}
c_{lmn}&=\psi_{0,lmn} + \eta_{lmn}\left[2 \tanh{\left( \frac{\epsilon_{lmn}-\mu}{2k_{B}T}\right)}\right]^{-1/2}.
\label{thermalnoise}
\end{align}
Here $\epsilon_{lmn} = \epsilon_{l} + \epsilon_{m} + \epsilon_{n}$ 
are the energies of the oscillator modes with quantum numbers $l,m,n$. $T$ is the temperature of the cloud and $\psi_{0,lmn}$ are the corresponding expansion coefficients of the GP ground state. The $\eta_{lmn}$ are complex Gaussian noises obeying $\overline{\eta_{lmn}^*\eta_{l^\prime m^\prime n^\prime}}=\delta_{l l^\prime}\delta_{m m^\prime} \delta_{n n^\prime}$.
Although the temperature in the Bosenova experiment was measured to be $3$ nK, in \fref{thermal_collapse} we present our results for a few different temperatures: $T=0$,  $T=3$~nK,  $T=5.3$~nK and  $T=8$~nK. We plot the final $2$~ms of simulated time. This corresponds to the collapse stage and exceeds the time of $\sim8$~ms for which we can employ sufficiently many modes to ensure a reliable simulation. Nonetheless we would like to draw some qualitative conclusions from these simulations of ``BEC collapse in a restricted mode space''. Firstly, if the collapse time was deduced from the condensate atom number alone, which is shown in \fref{thermal_collapse}~(a), it appears to be shorter for increased temperature. The reduction by $\sim 0.75$~ms for the experimental temperature of $3$~nK is however by far not enough to reach agreement with the experimental collapse time of $(6\pm1)$~ms. Secondly, the reduction in condensate atom number just before collapse, compared to the GP dynamics, results from stimulated transitions to uncondensed modes rather than an increased total atom loss, which can be deduced from an inspection of $\sub{N}{unc}$ and  $\sub{N}{tot}$. These \emph{qualitative} features are independent of the value of $\sub{E}{cut}$ used in the simulations. However we point out that the \emph{quantitative} details of the evolution of condensed and uncondensed fractions for the times presented in \fref{thermal_collapse} depend on $\sub{E}{cut}$. \fref{thermal_collapse}~(b) shows that the acceleration of collapse is much smaller, if only the total atom number is taken into account. 

In the experiment \cite{jila:nova}, the measured atom number was deduced from counting atoms in the central region of the trap. The increased population of the uncondensed modes just before collapse occupies the same spatial region as the condensate. Since the experimental method did not distinguish between condensed and uncondensed atoms, we conclude that the experiment probably did not capture the above described temperature effect. We note that, for $T\neq0$, the evolution of the \emph{total} number of atoms is only marginally changed compared to the GP results.

\section{HFB vs.~Wigner: Collapse of a spherical BEC}
\label{spherical}
In this section we compare the two different quantum field models of BECs used in this paper. Both rely on approximations to achieve a numerically tractable description of the quantum evolution. The formalism of each method and the approximations involved differ greatly, as outlined in \sref{models}. The \emph{quantitative} agreement between the evolution of the uncondensed fraction in both methods that we present in this section gives thus a strong indication of the validity of our results.

As has been described in Ref.~\cite{wuester:nova}, Hartree-Fock Bogoliubov simulations are not feasible in the case of the  cylindrically symmetric experimental situation, as the correlation functions $G_{A}$ and $G_{N}$ become then five dimensional. Therefore in our earlier work~\cite{wuester:nova}, we have used the HFB to investigate a collapse in a spherically symmetric trap. For the same reasons, we compare here the TWA and HFB methods for a range of temperatures of the initial thermal cloud in a \emph{spherical} geometry. Finite temperature is taken into account in the HFB approach by using correlation functions corresponding to a thermal population of oscillator states initially:
\begin{align}
\gnx&=\sum_{lmn}\frac{1}{\exp( {\frac{\epsilon_{lmn}-\mu}{k_{B}T}} ) -1}\varphi_{lmn}^{*}(\bv{x'})\varphi_{lmn}(\bv{x}),
\\
\gax&=0.
\label{temperature_hfb}
\end{align}
Here $\varphi_{lmn}(\bv{x})\equiv \varphi_{l}(x)\varphi_{m}(y)\varphi_{n}(z)$. For the finite temperature TWA simulations we have used $\sub{E}{cut}=40$. 

\begin{figure}[htb]
\centering
\epsfig{file={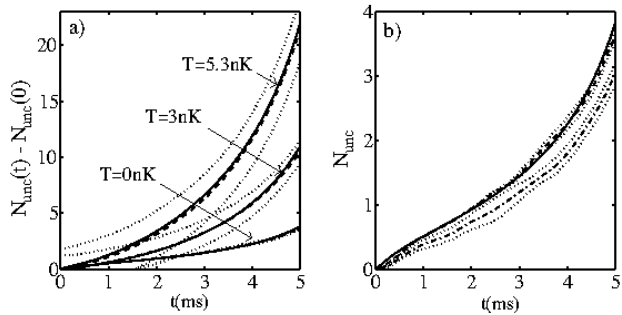},width=9.5cm} 
\caption{Increase in uncondensed atom number $\sub{N}{unc}(t)-\sub{N}{unc}(0)$ during the first stages of a spherical collapse with $\sub{a}{collapse}=-12 \sub{a}{0}$ for the TWA and HFB. (a) The TWA results with dynamical noise (dashed) are in excellent agreement with the HFB result (solid) for $T=0,3,5.3$~nK. Dotted lines indicate the sampling error.  Numerical parameters are given in the footnote \cite{fig4 numerics}. (b) Close-up of the result for $T=0$. As mentioned in \sref{cigar}, the TWA with dynamical noise (dashed) agrees better with the HFB (solid) than the result of the TWA with initial noise terms only (dot-dashed).
}
\label{wigner_hfb}
\end{figure}

We find excellent agreement between the uncondensed atom numbers predicted by HFB and TWA for the initial $5$ ms of the collapse~\cite{footnote:simperiod} and a range of temperatures, as shown in \fref{wigner_hfb}. The higher the temperature, the larger the initial uncondensed population $\sub{N}{unc}(0)$, which causes more stimulated transitions to the uncondensed fraction. As reported in~\cite{janne:wigner} an increase in temperature of the initial state also requires more trajectories for the sampling error to be satisfactorily small. 

The TWA and HFB in the spherical case both qualitatively confirm the results presented in \sref{thermal} for cylindrical geometry: higher temperature thermal clouds yield an increased creation of uncondensed particles just before collapse. This appears like an accelerated collapse on the curve for $\sub{N}{cond}$. In contrast, inspection of the total atom number shows almost no acceleration. 

{\noi \it Validity timescale:}
For the period we consider here, the approximations involved in both methods are justified and therefore a comparison is useful. It is known that the truncation in the TWA and the factorization of correlation functions in the HFB are valid for short times only, but both methods suffer from a lack of quantitative knowledge about this timescale in the general case. For the situation of a BEC in an optical lattice within the tight binding (Bose-Hubbard) regime, the validity timescales for TWA and HFB have been shown to coincide. They are given by  $t\ll J/U$, where $J$ and $U$ are the Bose-Hubbard hopping strength and on-site interaction respectively~\cite{polkovnikov:timescale,rey:timescale}. 

{\noi \it Numerical performance:}
We find that both methods for the quantum field treatment of Bose gases, HFB and TWA, agree in a direct comparison of the uncondensed atom number evolution during a BEC collapse. The TWA is advantageous for spatially asymmetric problems, as the increasing dimensionality of the correlation functions renders the HFB method numerically intractable. However, in a spherically symmetric case the HFB is advantageous. This is because the correlation functions $G_{N}$ and $G_{A}$ are only three dimensional due to the spatial symmetry. Meanwhile the truncated Wigner evolution has to be always modelled in full three dimensions as the quantum fluctuations cannot be assumed to be spherically symmetric. The quantum evolution in the HFB approach is obtained with a single solution of equations~(\ref{atom_eqn}), (\ref{ga_eqn}) and (\ref{gn_eqn}), compared to averaging over many realizations of \eref{sgpe} in the TWA. As a result our simulations showed that for the spherical case the HFB requires vastly shorter CPU times~\cite{footnote:runtimes}.

\section{Conclusions}
\label{conclusions}
Our three dimensional simulations of the Bosenova experiment 
on collapsing BECs \cite{jila:nova} have shown a moderate 
acceleration of the collapse if an initial thermal cloud
is taken into account, although the effect is not large
enough to quantitatively account for the discrepancy between
experimental and theoretical collapse times. 
The predictions of Hartree-Fock Bogoliubov and truncated Wigner theories for collapse in a spherically symmetric case 
agree very well with each other. However in a cigar shaped trap, where only the truncated Wigner method is feasible, the theory disagrees with the experiment. The origin of this discrepancy is unknown. 

Close to a Feshbach resonance molecular effects can become important. However during the sequence of the Bosenova experiment, the magnetic field stays clear of the resonance value \cite{jila:nova}. Hence molecules play a minor role in this experiment, as already argued before \cite{calz:hu,wuester:nova}. Also a possible breakdown of the s-wave approximation and strong effects due to inelastic collisions do not occur \emph{prior} to the actual collapse and hence cannot affect the collapse time. As other options are ruled out, we conjecture that the collapse time discrepancy arises from quantum correlations not captured by our descriptions. Although a Gaussian initial quantum state is physically reasonable, high order correlations can be created by the interactions and could become significant during collapse. These are not captured by the methods employed here.

To understand the discrepancy better, it would be desirable to conduct an experiment with the aim of measuring the collapse times and correlation functions for a larger range of scenarios and with even higher precision.

\begin{appendix}

\section{Projected Gross-Pitaevskii equation in the energy basis}
\label{pgpe_appendix}
Using the expansion (\ref{osciexpansion}) the stochastic equation (\ref{sgpe}) becomes:
\begin{align}
d c_{lmn}
&=-\frac{i}{\hbar}\left(\epsilon_{l} +\epsilon_{m}+\epsilon_{n} +\sub{U}{0} F_{lmn} \right)dt
\CR
&-\frac{K_3}{2}G_{lmn}dt + \sqrt{\frac{3 K_3}{2}} dH_{lmn}.
\label{sgpe_osci}
\end{align}
The $F$, $G$ and $dH$ are overlap integrals defined by:
\begin{align}
F_{lmn}&=\intas \varphi_{l}^{*}(x)\varphi_{m}^{*}(y)\varphi_{n}^{*}(z)|\phi(\mathbf{x})|^2\phi(\mathbf{x}),
\label{overlapp1}
 \\
G_{lmn}&=\intas \varphi_{l}^{*}(x)\varphi_{m}^{*}(y)\varphi_{n}^{*}(z)|\phi(\mathbf{x})|^4\phi(\mathbf{x}),
\label{overlapp2}
  \\
dH_{lmn}&=\intas \varphi_{l}^{*}(x)\varphi_{m}^{*}(y)\varphi_{n}^{*}(z)|\phi(\mathbf{x})|^2 d\xi(\bv{x}),
\label{overlapp3}
\\
d\xi(\bv{x})&=\sum_{ \{l,m,n\} \in {\cal C} }d\xi_{lmn}(t)\varphi_{l}(x)\varphi_{m}(y)\varphi_{n}(z).
\label{dynnoise}
\end{align}
$d\xi_{lmn}(t)$ in \eref{dynnoise} are complex Gaussian noises that fulfill $\overline{d\xi_{lmn}(t)d\xi_{l'm'n'}(t')} = 0$ and $\overline{d\xi_{lmn}^{*}(t)d\xi_{l'm'n'}(t)} = \delta_{l l^\prime}\delta_{m m^\prime}\delta_{n n^\prime}dt$.

It has been outlined in Refs.~\cite{matthewandblair:pgpe,dion:hermites} how these integrals can separately be exactly computed on an appropriately chosen non-equidistant spatial grid. Different spatial grids would however be necessary for the exact solution of integrals involving different powers of the wavefunction. To remain computationally efficient we have chosen a grid which allows the exact calculation of Eqns.~(\ref{overlapp1}) and (\ref{overlapp3}). We have checked that our results are invariant under a variation of the grid used for evaluation of the integrals.

\section{HFB equations for correlation functions}
\label{hfb_appendix}

In the time-dependent Hartree-Fock-Bogoliubov approach we have to supplement \eref{atom_eqn} by evolution equations for $G_{N}$ and $G_{A}$. 
These are obtained by deriving the Heisenberg equations for the operators $\hat{\chi}^{\dagger}(x')\hat{\chi}(x)$ and $\hat{\chi}(x')\hat{\chi}(x)$ respectively, and factorizing operator products as described in~\cite{wuester:nova}. This procedure yields:
\begin{align}
\nonumber
&i \hbar  \frac{\partial \gnx}{\partial t}=
\langle \big[\hat{\chi}^{\dagger}(x')\hat{\chi}(x), \hat{H}\big]\rangle=
\nnl
&\big( \hamax - \hamaxs \big) \gnx +2 \sub{U}{0} \big( \mods{\phiax}  - \mods{\phiaxs}
\nnl
&+ \gnxx - G_{N}\left( \mathbf{x'},\mathbf{x'} \right)  \big) \gnx 
\nnl
 &+ \sub{U}{0} \big( \gaxx \gaxstar- G_{A}^{*}\left( \mathbf{x'},\mathbf{x'} \right)\gax \big)
\nl
&+\sub{U}{0} \big(\phiax^{2}\gaxstar -\phicaxs^{2} \gax \big),
\label{gn_eqn}
\end{align}
\begin{align}
\nonumber
&i \hbar  \frac{\partial \gax}{\partial t}=
\langle \big[\hat{\chi}(x')\hat{\chi}(x) , \hat{H}\big]\rangle=
\nnl
&\big( \hamax + \hamaxs \big) \gax +2 \sub{U}{0} \big( \mods{\phiax}  + \mods{\phiaxs} 
\nnl
&+ \gnxx +G_{N}\left( \mathbf{x'},\mathbf{x'} \right) \big) \gax 
\nnl
&+ \sub{U}{0} \big(  \phiax^{2} \gnxstar  +\phiaxs^{2} \gnx 
\nnl
&+ \gaxx  \gnxstar   + G_{A}\left( \mathbf{x'},\mathbf{x'}\right) \gnx \big) 
\nl
&+ \sub{U}{0} \big(\phiax^{2} + \gaxx \big)\deltf.
\label{ga_eqn}
\end{align}
\end{appendix}
\acknowledgments
BJDW and SW are grateful for the very kind hospitality of the Quantum-Atom
optics theory group at the University of Queensland. This research was supported by the Australian Research Council under the Centre of Excellence for Quantum-Atom Optics and by an award under the Merit Allocation Scheme of the National Facility of the Australian Partnership for Advanced Computing.


\begin{thebibliography}{51}
\expandafter\ifx\csname natexlab\endcsname\relax\def\natexlab#1{#1}\fi
\expandafter\ifx\csname bibnamefont\endcsname\relax
  \def\bibnamefont#1{#1}\fi
\expandafter\ifx\csname bibfnamefont\endcsname\relax
  \def\bibfnamefont#1{#1}\fi
\expandafter\ifx\csname citenamefont\endcsname\relax
  \def\citenamefont#1{#1}\fi
\expandafter\ifx\csname url\endcsname\relax
  \def\url#1{\texttt{#1}}\fi
\expandafter\ifx\csname urlprefix\endcsname\relax\def\urlprefix{URL }\fi
\providecommand{\bibinfo}[2]{#2}
\providecommand{\eprint}[2][]{\url{#2}}

\bibitem[{\citenamefont{Schellekens et~al.}(2005)\citenamefont{Schellekens,
  Hoppler, Perrin, Gomes, Boiron, Aspect, and Westbrook}}]{Schellekens2005}
\bibinfo{author}{\bibfnamefont{M.}~\bibnamefont{Schellekens}},
  \bibinfo{author}{\bibfnamefont{R.}~\bibnamefont{Hoppler}},
  \bibinfo{author}{\bibfnamefont{A.}~\bibnamefont{Perrin}},
  \bibinfo{author}{\bibfnamefont{J.~V.} \bibnamefont{Gomes}},
  \bibinfo{author}{\bibfnamefont{D.}~\bibnamefont{Boiron}},
  \bibinfo{author}{\bibfnamefont{A.~A.} \bibnamefont{Aspect}},
  \bibnamefont{and} \bibinfo{author}{\bibfnamefont{C.~I.}
  \bibnamefont{Westbrook}}, \bibinfo{journal}{Science}
  \textbf{\bibinfo{volume}{310}}, \bibinfo{pages}{648} (\bibinfo{year}{2005}).

\bibitem[{\citenamefont{Chuu et~al.}(2005)\citenamefont{Chuu, Schreck, Meyrath,
  Hanssen, Price, and Raizen}}]{Chuu2005}
\bibinfo{author}{\bibfnamefont{C.-S.} \bibnamefont{Chuu}},
  \bibinfo{author}{\bibfnamefont{F.}~\bibnamefont{Schreck}},
  \bibinfo{author}{\bibfnamefont{T.~P.} \bibnamefont{Meyrath}},
  \bibinfo{author}{\bibfnamefont{J.~L.} \bibnamefont{Hanssen}},
  \bibinfo{author}{\bibfnamefont{G.~N.} \bibnamefont{Price}}, \bibnamefont{and}
  \bibinfo{author}{\bibfnamefont{M.~G.} \bibnamefont{Raizen}},
  \bibinfo{journal}{Phys. Rev. Lett.} \textbf{\bibinfo{volume}{95}},
  \bibinfo{pages}{260403} (\bibinfo{year}{2005}).

\bibitem[{\citenamefont{F\"{o}lling et~al.}(2005)\citenamefont{F\"{o}lling,
  Gerbler, Widera, Mandel, Gericke, and Bloch}}]{Folling2005}
\bibinfo{author}{\bibfnamefont{S.}~\bibnamefont{F\"{o}lling}},
  \bibinfo{author}{\bibfnamefont{F.}~\bibnamefont{Gerbler}},
  \bibinfo{author}{\bibfnamefont{A.}~\bibnamefont{Widera}},
  \bibinfo{author}{\bibfnamefont{O.}~\bibnamefont{Mandel}},
  \bibinfo{author}{\bibfnamefont{T.}~\bibnamefont{Gericke}}, \bibnamefont{and}
  \bibinfo{author}{\bibfnamefont{I.}~\bibnamefont{Bloch}},
  \bibinfo{journal}{Nature} \textbf{\bibinfo{volume}{434}},
  \bibinfo{pages}{481} (\bibinfo{year}{2005}).

\bibitem[{\citenamefont{\"{O}ttl et~al.}(2005)\citenamefont{\"{O}ttl, Ritter,
  K\"{o}hl, and Esslinger}}]{Ottl2005}
\bibinfo{author}{\bibfnamefont{A.}~\bibnamefont{\"{O}ttl}},
  \bibinfo{author}{\bibfnamefont{S.}~\bibnamefont{Ritter}},
  \bibinfo{author}{\bibfnamefont{M.}~\bibnamefont{K\"{o}hl}}, \bibnamefont{and}
  \bibinfo{author}{\bibfnamefont{T.}~\bibnamefont{Esslinger}},
  \bibinfo{journal}{Phys. Rev. Lett.} \textbf{\bibinfo{volume}{95}},
  \bibinfo{pages}{090404} (\bibinfo{year}{2005}).

\bibitem[{\citenamefont{Gardiner and Zoller}(2004)}]{book:qn}
\bibinfo{author}{\bibfnamefont{C.~W.} \bibnamefont{Gardiner}} \bibnamefont{and}
  \bibinfo{author}{\bibfnamefont{P.}~\bibnamefont{Zoller}},
  \emph{\bibinfo{title}{Quantum Noise}} (\bibinfo{publisher}{Springer-Verlag,
  Berlin Heidelberg,}, \bibinfo{year}{2004}).

\bibitem[{\citenamefont{Morgan et~al.}(2003)\citenamefont{Morgan, Rusch,
  Hutchinson, and Burnett}}]{morgan:hfb}
\bibinfo{author}{\bibfnamefont{S.~A.} \bibnamefont{Morgan}},
  \bibinfo{author}{\bibfnamefont{M.}~\bibnamefont{Rusch}},
  \bibinfo{author}{\bibfnamefont{D.~A.~W.} \bibnamefont{Hutchinson}},
  \bibnamefont{and} \bibinfo{author}{\bibfnamefont{K.}~\bibnamefont{Burnett}},
  \bibinfo{journal}{Phys. Rev. Lett.} \textbf{\bibinfo{volume}{91}},
  \bibinfo{pages}{250403} (\bibinfo{year}{2003}).

\bibitem[{\citenamefont{Savage et~al.}(2003)\citenamefont{Savage, Robins, and
  Hope}}]{savage:coll}
\bibinfo{author}{\bibfnamefont{C.~M.} \bibnamefont{Savage}},
  \bibinfo{author}{\bibfnamefont{N.~P.} \bibnamefont{Robins}},
  \bibnamefont{and} \bibinfo{author}{\bibfnamefont{J.~J.} \bibnamefont{Hope}},
  \bibinfo{journal}{Phys. Rev. A} \textbf{\bibinfo{volume}{67}},
  \bibinfo{pages}{014304} (\bibinfo{year}{2003}).

\bibitem[{\citenamefont{W{\"u}ster et~al.}(2005)\citenamefont{W{\"u}ster, Hope,
  and Savage}}]{wuester:nova}
\bibinfo{author}{\bibfnamefont{S.}~\bibnamefont{W{\"u}ster}},
  \bibinfo{author}{\bibfnamefont{J.~J.} \bibnamefont{Hope}}, \bibnamefont{and}
  \bibinfo{author}{\bibfnamefont{C.~M.} \bibnamefont{Savage}},
  \bibinfo{journal}{Phys. Rev. A} \textbf{\bibinfo{volume}{71}},
  \bibinfo{pages}{033604} (\bibinfo{year}{2005}).

\bibitem[{\citenamefont{Donley et~al.}(2001)\citenamefont{Donley, Claussen,
  Cornish, Roberts, Cornell, and Wieman}}]{jila:nova}
\bibinfo{author}{\bibfnamefont{E.~A.} \bibnamefont{Donley}},
  \bibinfo{author}{\bibfnamefont{N.~R.} \bibnamefont{Claussen}},
  \bibinfo{author}{\bibfnamefont{S.~L.} \bibnamefont{Cornish}},
  \bibinfo{author}{\bibfnamefont{J.~L.} \bibnamefont{Roberts}},
  \bibinfo{author}{\bibfnamefont{E.~A.} \bibnamefont{Cornell}},
  \bibnamefont{and} \bibinfo{author}{\bibfnamefont{C.~E.}
  \bibnamefont{Wieman}}, \bibinfo{journal}{Nature}
  \textbf{\bibinfo{volume}{412}}, \bibinfo{pages}{295} (\bibinfo{year}{2001}).

\bibitem[{\citenamefont{Santos and Shlyapnikov}(2002)}]{shlyapnikov:coll}
\bibinfo{author}{\bibfnamefont{L.}~\bibnamefont{Santos}} \bibnamefont{and}
  \bibinfo{author}{\bibfnamefont{G.~V.} \bibnamefont{Shlyapnikov}},
  \bibinfo{journal}{Phys. Rev. A} \textbf{\bibinfo{volume}{66}},
  \bibinfo{pages}{011602(R)} (\bibinfo{year}{2002}).

\bibitem[{\citenamefont{Adhikari}(2002)}]{adhikari:coll}
\bibinfo{author}{\bibfnamefont{S.~K.} \bibnamefont{Adhikari}},
  \bibinfo{journal}{Physics Letters A} \textbf{\bibinfo{volume}{296}},
  \bibinfo{pages}{145} (\bibinfo{year}{2002}).

\bibitem[{\citenamefont{Adhikari}(2004)}]{adhikari:jets}
\bibinfo{author}{\bibfnamefont{S.~K.} \bibnamefont{Adhikari}},
  \bibinfo{journal}{J. Phys. B} \textbf{\bibinfo{volume}{37}},
  \bibinfo{pages}{1185} (\bibinfo{year}{2004}).

\bibitem[{\citenamefont{Adhikari}(2005)}]{adhikari:lattice}
\bibinfo{author}{\bibfnamefont{S.~K.} \bibnamefont{Adhikari}},
  \bibinfo{journal}{Phys. Rev. A} \textbf{\bibinfo{volume}{71}},
  \bibinfo{pages}{053603} (\bibinfo{year}{2005}).

\bibitem[{\citenamefont{Saito and Ueda}(2002)}]{saito_ueda:coll}
\bibinfo{author}{\bibfnamefont{H.}~\bibnamefont{Saito}} \bibnamefont{and}
  \bibinfo{author}{\bibfnamefont{M.}~\bibnamefont{Ueda}},
  \bibinfo{journal}{Phys. Rev. A} \textbf{\bibinfo{volume}{65}},
  \bibinfo{pages}{033624} (\bibinfo{year}{2002}).

\bibitem[{\citenamefont{Saito and Ueda}(2000)}]{saito_ueda:intermitt}
\bibinfo{author}{\bibfnamefont{H.}~\bibnamefont{Saito}} \bibnamefont{and}
  \bibinfo{author}{\bibfnamefont{M.}~\bibnamefont{Ueda}},
  \bibinfo{journal}{Phys. Rev. Lett.} \textbf{\bibinfo{volume}{86}},
  \bibinfo{pages}{1406} (\bibinfo{year}{2000}).

\bibitem[{\citenamefont{Saito and Ueda}(2001)}]{saito_ueda:powerlaw}
\bibinfo{author}{\bibfnamefont{H.}~\bibnamefont{Saito}} \bibnamefont{and}
  \bibinfo{author}{\bibfnamefont{M.}~\bibnamefont{Ueda}},
  \bibinfo{journal}{Phys. Rev. A} \textbf{\bibinfo{volume}{63}},
  \bibinfo{pages}{043601} (\bibinfo{year}{2001}).

\bibitem[{\citenamefont{Ueda and Saito}(2003)}]{saito_ueda:picture}
\bibinfo{author}{\bibfnamefont{M.}~\bibnamefont{Ueda}} \bibnamefont{and}
  \bibinfo{author}{\bibfnamefont{H.}~\bibnamefont{Saito}}, \bibinfo{journal}{J.
  Phys. Soc. Jpn. Suppl. C} \textbf{\bibinfo{volume}{72}}, \bibinfo{pages}{127}
  (\bibinfo{year}{2003}).

\bibitem[{\citenamefont{Bao et~al.}(2004)\citenamefont{Bao, Jaksch, and
  Markowich}}]{bao:jets}
\bibinfo{author}{\bibfnamefont{W.}~\bibnamefont{Bao}},
  \bibinfo{author}{\bibfnamefont{D.}~\bibnamefont{Jaksch}}, \bibnamefont{and}
  \bibinfo{author}{\bibfnamefont{P.~A.} \bibnamefont{Markowich}},
  \bibinfo{journal}{J. Phys. B} \textbf{\bibinfo{volume}{37}},
  \bibinfo{pages}{329} (\bibinfo{year}{2004}).

\bibitem[{\citenamefont{Duine and Stoof}(2003)}]{duinestoof:feshbachdynamics}
\bibinfo{author}{\bibfnamefont{R.~A.} \bibnamefont{Duine}} \bibnamefont{and}
  \bibinfo{author}{\bibfnamefont{H.~T.~C.} \bibnamefont{Stoof}},
  \bibinfo{journal}{Phys. Rev. A} \textbf{\bibinfo{volume}{68}},
  \bibinfo{pages}{013602} (\bibinfo{year}{2003}).

\bibitem[{\citenamefont{Duine and Stoof}(2001)}]{duinestoof:evaporation}
\bibinfo{author}{\bibfnamefont{R.~A.} \bibnamefont{Duine}} \bibnamefont{and}
  \bibinfo{author}{\bibfnamefont{H.~T.~C.} \bibnamefont{Stoof}},
  \bibinfo{journal}{Phys. Rev. Lett.} \textbf{\bibinfo{volume}{86}},
  \bibinfo{pages}{2204} (\bibinfo{year}{2001}).

\bibitem[{\citenamefont{M\'etens et~al.}(2003)\citenamefont{M\'etens, Dewel,
  and Borckmans}}]{metens}
\bibinfo{author}{\bibfnamefont{S.}~\bibnamefont{M\'etens}},
  \bibinfo{author}{\bibfnamefont{G.}~\bibnamefont{Dewel}}, \bibnamefont{and}
  \bibinfo{author}{\bibfnamefont{P.}~\bibnamefont{Borckmans}},
  \bibinfo{journal}{Phys. Rev. A} \textbf{\bibinfo{volume}{68}},
  \bibinfo{pages}{045601} (\bibinfo{year}{2003}).

\bibitem[{\citenamefont{Calzetta and Hu}(2003)}]{calz:hu}
\bibinfo{author}{\bibfnamefont{E.~A.} \bibnamefont{Calzetta}} \bibnamefont{and}
  \bibinfo{author}{\bibfnamefont{B.~L.} \bibnamefont{Hu}},
  \bibinfo{journal}{Phys. Rev. A} \textbf{\bibinfo{volume}{68}},
  \bibinfo{pages}{043625} (\bibinfo{year}{2003}).

\bibitem[{\citenamefont{Milstein et~al.}(2003)\citenamefont{Milstein, Menotti,
  and Holland}}]{holland:burst}
\bibinfo{author}{\bibfnamefont{J.~N.} \bibnamefont{Milstein}},
  \bibinfo{author}{\bibfnamefont{C.}~\bibnamefont{Menotti}}, \bibnamefont{and}
  \bibinfo{author}{\bibfnamefont{M.~J.} \bibnamefont{Holland}},
  \bibinfo{journal}{New J. Phys.} \textbf{\bibinfo{volume}{5}},
  \bibinfo{pages}{52} (\bibinfo{year}{2003}).
  
  \bibitem[{\citenamefont{Cornish et~al.}(2006)\citenamefont{Cornish, Thompson,
  and Wieman}}]{jila:solitons}
\bibinfo{author}{\bibfnamefont{S.~L.} \bibnamefont{Cornish}},
  \bibinfo{author}{\bibfnamefont{S.~T.} \bibnamefont{Thompson}},
  \bibnamefont{and} \bibinfo{author}{\bibfnamefont{C.~E.}
  \bibnamefont{Wieman}}, \bibinfo{journal}{Phys. Rev. Lett.}
  \textbf{\bibinfo{volume}{96}}, \bibinfo{pages}{170401}
  (\bibinfo{year}{2006}).
  
  
\bibitem[{\citenamefont{Claussen et~al.}(2003)\citenamefont{Claussen,
  Kokkelmans, Thompson, Donley, Hodby, and Wieman}}]{jila:revision}
\bibinfo{author}{\bibfnamefont{N.~R.} \bibnamefont{Claussen}},
  \bibinfo{author}{\bibfnamefont{S.~J. J. M.~F.} \bibnamefont{Kokkelmans}},
  \bibinfo{author}{\bibfnamefont{S.~T.} \bibnamefont{Thompson}},
  \bibinfo{author}{\bibfnamefont{E.~A.} \bibnamefont{Donley}},
  \bibinfo{author}{\bibfnamefont{E.}~\bibnamefont{Hodby}}, \bibnamefont{and}
  \bibinfo{author}{\bibfnamefont{C.~E.} \bibnamefont{Wieman}},
  \bibinfo{journal}{Phys. Rev. A} \textbf{\bibinfo{volume}{67}},
  \bibinfo{pages}{060701(R)} (\bibinfo{year}{2003}).

\bibitem[{\citenamefont{Jack}(2002)}]{jack:loss}
\bibinfo{author}{\bibfnamefont{M.~W.} \bibnamefont{Jack}},
  \bibinfo{journal}{Phys. Rev. Lett.} \textbf{\bibinfo{volume}{89}},
  \bibinfo{pages}{140402} (\bibinfo{year}{2002}).

\bibitem[{\citenamefont{Steel et~al.}(1998)\citenamefont{Steel, Olsen, Plimak,
  Drummond, Tan, Collet, Walls, and Graham}}]{steel:wigner}
\bibinfo{author}{\bibfnamefont{M.~J.} \bibnamefont{Steel}},
  \bibinfo{author}{\bibfnamefont{M.~K.} \bibnamefont{Olsen}},
  \bibinfo{author}{\bibfnamefont{L.~I.} \bibnamefont{Plimak}},
  \bibinfo{author}{\bibfnamefont{P.~D.} \bibnamefont{Drummond}},
  \bibinfo{author}{\bibfnamefont{S.~M.} \bibnamefont{Tan}},
  \bibinfo{author}{\bibfnamefont{M.~J.} \bibnamefont{Collett}},
  \bibinfo{author}{\bibfnamefont{D.~F.} \bibnamefont{Walls}}, \bibnamefont{and}
  \bibinfo{author}{\bibfnamefont{R.}~\bibnamefont{Graham}},
  \bibinfo{journal}{Phys. Rev. A} \textbf{\bibinfo{volume}{58}},
  \bibinfo{pages}{4824} (\bibinfo{year}{1998}).

\bibitem[{\citenamefont{Sinatra et~al.}(2002)\citenamefont{Sinatra, Lobo, and
  Castin}}]{castin:validity}
\bibinfo{author}{\bibfnamefont{A.}~\bibnamefont{Sinatra}},
  \bibinfo{author}{\bibfnamefont{C.}~\bibnamefont{Lobo}}, \bibnamefont{and}
  \bibinfo{author}{\bibfnamefont{Y.}~\bibnamefont{Castin}},
  \bibinfo{journal}{J. Phys. B} \textbf{\bibinfo{volume}{35}},
  \bibinfo{pages}{3599} (\bibinfo{year}{2002}).

\bibitem[{\citenamefont{Norrie et~al.}(2006{\natexlab{a}})\citenamefont{Norrie,
  Ballagh, and Gardiner}}]{norrie:long}
\bibinfo{author}{\bibfnamefont{A.~A.} \bibnamefont{Norrie}},
  \bibinfo{author}{\bibfnamefont{R.~J.} \bibnamefont{Ballagh}},
  \bibnamefont{and} \bibinfo{author}{\bibfnamefont{C.~W.}
  \bibnamefont{Gardiner}}, \bibinfo{journal}{Phys. Rev. A}
  \textbf{\bibinfo{volume}{73}}, \bibinfo{pages}{043617}
  (\bibinfo{year}{2006}{\natexlab{a}}).

\bibitem[{\citenamefont{Deuar and
  Drummond}(2006{\natexlab{a}})}]{drummond:posp}
\bibinfo{author}{\bibfnamefont{P.}~\bibnamefont{Deuar}} \bibnamefont{and}
  \bibinfo{author}{\bibfnamefont{P.~D.} \bibnamefont{Drummond}},
  \bibinfo{journal}{J. Phys. A.: Math. Gen.} \textbf{\bibinfo{volume}{39}},
  \bibinfo{pages}{1163} (\bibinfo{year}{2006}{\natexlab{a}}).

\bibitem[{\citenamefont{Deuar and
  Drummond}(2006{\natexlab{b}})}]{drummond:gaugep}
\bibinfo{author}{\bibfnamefont{P.}~\bibnamefont{Deuar}} \bibnamefont{and}
  \bibinfo{author}{\bibfnamefont{P.~D.} \bibnamefont{Drummond}},
  \bibinfo{journal}{J. Phys. A.: Math. Gen.} \textbf{\bibinfo{volume}{39}},
  \bibinfo{pages}{2723} (\bibinfo{year}{2006}{\natexlab{b}}).

\bibitem[{\citenamefont{Norrie et~al.}(2006{\natexlab{b}})\citenamefont{Norrie,
  Ballagh, Gardiner, and Bradley}}]{ashton:loss}
\bibinfo{author}{\bibfnamefont{A.~A.} \bibnamefont{Norrie}},
  \bibinfo{author}{\bibfnamefont{R.~J.} \bibnamefont{Ballagh}},
  \bibinfo{author}{\bibfnamefont{C.~W.} \bibnamefont{Gardiner}},
  \bibnamefont{and} \bibinfo{author}{\bibfnamefont{A.~S.}
  \bibnamefont{Bradley}}, \bibinfo{journal}{Phys. Rev. A}
  \textbf{\bibinfo{volume}{73}}, \bibinfo{pages}{043618}
  (\bibinfo{year}{2006}{\natexlab{b}}).

\bibitem[{foo({\natexlab{a}})}]{foonote:simplest}
\bibinfo{note}{Most of the extra terms can be neglected in a regime where the
  necessary truncation is valid \cite{ashton:loss}.}

\bibitem[{\citenamefont{D\c{a}browska-W{\"u}ster
  et~al.}(2006)\citenamefont{D\c{a}browska-W{\"u}ster, W{\"u}ster, Bradley,
  Davis, and Ostrovskaya}}]{beatka:train}
\bibinfo{author}{\bibfnamefont{B.~J.} \bibnamefont{D\c{a}browska-W{\"u}ster}},
  \bibinfo{author}{\bibfnamefont{S.}~\bibnamefont{W{\"u}ster}},
  \bibinfo{author}{\bibfnamefont{A.~S.} \bibnamefont{Bradley}},
  \bibinfo{author}{\bibfnamefont{M.~J.} \bibnamefont{Davis}}, \bibnamefont{and}
  \bibinfo{author}{\bibfnamefont{E.~A.} \bibnamefont{Ostrovskaya}}
  (\bibinfo{year}{2006}), \eprint{cond-mat/0607332}.

\bibitem[{\citenamefont{Dion and Canc{\`e}s}(2003)}]{dion:hermites}
\bibinfo{author}{\bibfnamefont{C.~M.} \bibnamefont{Dion}} \bibnamefont{and}
  \bibinfo{author}{\bibfnamefont{E.}~\bibnamefont{Canc{\`e}s}},
  \bibinfo{journal}{Phys. Rev. E} \textbf{\bibinfo{volume}{67}},
  \bibinfo{pages}{046706} (\bibinfo{year}{2003}).

\bibitem[{\citenamefont{Blakie and Davis}(2005)}]{matthewandblair:pgpe}
\bibinfo{author}{\bibfnamefont{P.~B.} \bibnamefont{Blakie}} \bibnamefont{and}
  \bibinfo{author}{\bibfnamefont{M.~J.} \bibnamefont{Davis}},
  \bibinfo{journal}{Phys. Rev. A} \textbf{\bibinfo{volume}{72}},
  \bibinfo{pages}{063608} (\bibinfo{year}{2005}).

\bibitem[{\citenamefont{Davis and Blakie}(2006)}]{matthewandblair:pgpeprl}
\bibinfo{author}{\bibfnamefont{M.~J.} \bibnamefont{Davis}} \bibnamefont{and}
  \bibinfo{author}{\bibfnamefont{P.~B.} \bibnamefont{Blakie}},
  \bibinfo{journal}{Phys. Rev. Lett.} \textbf{\bibinfo{volume}{96}},
  \bibinfo{pages}{060404} (\bibinfo{year}{2006}).

\bibitem[{foo({\natexlab{b}})}]{footnote:trajectories}
\bibinfo{note}{For the results presented in this paper we used from 50 to 400
  trajectories. Higher temperatures required more trajectories, as did the
  dynamical noise evolution compared to the initial noise only.}

\bibitem[{\citenamefont{Penrose and Onsager}(1956)}]{penrose:condensate}
\bibinfo{author}{\bibfnamefont{O.}~\bibnamefont{Penrose}} \bibnamefont{and}
  \bibinfo{author}{\bibfnamefont{L.}~\bibnamefont{Onsager}},
  \bibinfo{journal}{Phys. Rev.} \textbf{\bibinfo{volume}{104}},
  \bibinfo{pages}{576} (\bibinfo{year}{1956}).

\bibitem[{\citenamefont{G{\'o}ral et~al.}(2002)\citenamefont{G{\'o}ral, Gajda,
  and Rz{\c{a}}{\.z}ewski}}]{gajda:classfield}
\bibinfo{author}{\bibfnamefont{K.}~\bibnamefont{G{\'o}ral}},
  \bibinfo{author}{\bibfnamefont{M.}~\bibnamefont{Gajda}}, \bibnamefont{and}
  \bibinfo{author}{\bibfnamefont{K.}~\bibnamefont{Rz{\c{a}}{\.z}ewski}},
  \bibinfo{journal}{Phys. Rev. A} \textbf{\bibinfo{volume}{66}},
  \bibinfo{pages}{051602(R)} (\bibinfo{year}{2002}).

\bibitem[{\citenamefont{Norrie et~al.}(2005)\citenamefont{Norrie, Ballagh, and
  Gardiner}}]{norrie:prl}
\bibinfo{author}{\bibfnamefont{A.~A.} \bibnamefont{Norrie}},
  \bibinfo{author}{\bibfnamefont{R.~J.} \bibnamefont{Ballagh}},
  \bibnamefont{and} \bibinfo{author}{\bibfnamefont{C.~W.}
  \bibnamefont{Gardiner}}, \bibinfo{journal}{Phys. Rev. Lett.}
  \textbf{\bibinfo{volume}{94}}, \bibinfo{pages}{040401}
  (\bibinfo{year}{2005}).

\bibitem[{\citenamefont{Blaizot and Ripka}(1986)}]{book:blaizot}
\bibinfo{author}{\bibfnamefont{J.-P.} \bibnamefont{Blaizot}} \bibnamefont{and}
  \bibinfo{author}{\bibfnamefont{G.}~\bibnamefont{Ripka}},
  \emph{\bibinfo{title}{Quantum theory of finite systems}}
  (\bibinfo{publisher}{MIT Press}, \bibinfo{year}{1986}).

\bibitem[{\citenamefont{Kokkelmans et~al.}(2002)\citenamefont{Kokkelmans,
  Milstein, Chiofalo, Walser, and Holland}}]{holland:renorm}
\bibinfo{author}{\bibfnamefont{S.~J. J. M.~F.} \bibnamefont{Kokkelmans}},
  \bibinfo{author}{\bibfnamefont{J.~N.} \bibnamefont{Milstein}},
  \bibinfo{author}{\bibfnamefont{M.~L.} \bibnamefont{Chiofalo}},
  \bibinfo{author}{\bibfnamefont{R.}~\bibnamefont{Walser}}, \bibnamefont{and}
  \bibinfo{author}{\bibfnamefont{M.~J.} \bibnamefont{Holland}},
  \bibinfo{journal}{Phys. Rev. A} \textbf{\bibinfo{volume}{65}},
  \bibinfo{pages}{053617} (\bibinfo{year}{2002}).


\bibitem[{foo({\natexlab{d}})}]{footnote:notfft}
\bibinfo{note}{Note that in the oscillator basis we cannot make use of the FFT
  algorithm, and hence the tractable mode numbers are lower than for position
  grid methods.}

\bibitem[{foo({\natexlab{e}})}]{footnote:samplingerror}
\bibinfo{note}{We determine the sampling error from the difference between the
  average of all and $75\%$ of the trajectories.}

\bibitem[{foo({\natexlab{f}})}]{footnote:simperiod}
\bibinfo{note}{For this period the evolution can be obtained with $\sim5\times
  10^{4}$ modes, allowing multi-trajectory solutions.}

\bibitem[{\citenamefont{Isella and Ruostekoski}(2005)}]{janne:wigner}
\bibinfo{author}{\bibfnamefont{L.}~\bibnamefont{Isella}} \bibnamefont{and}
  \bibinfo{author}{\bibfnamefont{J.}~\bibnamefont{Ruostekoski}},
  \bibinfo{journal}{Phys. Rev. A} \textbf{\bibinfo{volume}{72}},
  \bibinfo{pages}{011601(R)} (\bibinfo{year}{2005}).

\bibitem[{\citenamefont{Polkovnikov}(2003)}]{polkovnikov:timescale}
\bibinfo{author}{\bibfnamefont{A.}~\bibnamefont{Polkovnikov}},
  \bibinfo{journal}{Phys. Rev. A} \textbf{\bibinfo{volume}{68}},
  \bibinfo{pages}{053604} (\bibinfo{year}{2003}).

\bibitem[{\citenamefont{Rey et~al.}(2004)\citenamefont{Rey, Hu, Calzetta,
  Roura, and Clark}}]{rey:timescale}
\bibinfo{author}{\bibfnamefont{A.~M.} \bibnamefont{Rey}},
  \bibinfo{author}{\bibfnamefont{B.~L.} \bibnamefont{Hu}},
  \bibinfo{author}{\bibfnamefont{E.}~\bibnamefont{Calzetta}},
  \bibinfo{author}{\bibfnamefont{A.}~\bibnamefont{Roura}}, \bibnamefont{and}
  \bibinfo{author}{\bibfnamefont{C.~W.} \bibnamefont{Clark}},
  \bibinfo{journal}{Phys. Rev. A} \textbf{\bibinfo{volume}{69}},
  \bibinfo{pages}{033610} (\bibinfo{year}{2004}).
  
\bibitem[{foo({\natexlab{c}})}]{fig4 numerics}
\bibinfo{note}{For the TWA simulations $\sub{E}{cut}=50$ for $T=0$ and $\sub{E}{cut}=40$ for $T \neq 0$. For the HFB simulations $128 \times 128$ grids were used with lengths of $30, 50$ and $76$ $\mu$m for $T=0, 3$ and $5.3$ nK respectively.}

\bibitem[{foo({\natexlab{g}})}]{footnote:runtimes}
\bibinfo{note}{As example: For the result shown in \fref{wigner_hfb} (b), the
  HFB took 26 CPU hours (using parallel computation), while the 200
  trajectories of the TWA took 450 CPU hours.}

\end{thebibliography}
\end{document}